\documentstyle[epsfig,prl,aps]{revtex}
\begin{document}
\draft
\twocolumn[\hsize\textwidth\columnwidth\hsize\csname
@twocolumnfalse\endcsname
\title{ The algebraic structure of a cosmological term in spherically
symmetric solutions
}
\author{I. G. Dymnikova
}
\address{Institute of Mathematics, Informatics and Physics,
         University of Olsztyn,
         10-561 Olsztyn, Poland
}
\maketitle

\begin{abstract} 
We propose to describe the dynamics of a cosmological
term in the spherically symmetric case by an $r$-dependent second rank 
symmetric 
tensor $\Lambda_{\mu\nu}$ invariant under boosts in the radial direction. 
This proposal is based on the Petrov classification scheme and Einstein
field equations in the spherically symmetric case.
The inflationary equation of state $p=-\rho$ is satisfied by the radial
pressure, ${p}_r^{\Lambda}=-{\rho}^{\Lambda}$.
The tangential pressure ${p}_{\perp}^{\Lambda}$ is calculated from the 
conservation equation $\Lambda^{\mu}_{\nu;\mu}=0$.

\end{abstract}


\vspace{0.2cm}
]
%\twocolumn

Developments in particles and quantum field theory, as well as the
confrontation of models with observations in cosmology \cite{bahcall}, 
compellingly favour treating the cosmological constant $\Lambda$ 
as a dynamical quantity.

The Einstein equations with a cosmological term read
$$
G_{\mu\nu}+\Lambda g_{\mu\nu}=-8\pi G T_{\mu\nu}\eqno(1)
$$
where $G_{\mu\nu}$ is the Einstein tensor, 
$T_{\mu\nu}$ is the stress-energy tensor
of a matter, and $\Lambda$ is the cosmological constant.
In the absence of  matter described by $T_{\mu\nu}$, $\Lambda$
must be constant, since the Bianchi identities guarantee vanishing
covariant divergence of the Einstein tensor, $G^{\mu\nu}_{~~;\nu}=0$.

 In quantum field theory, the vacuum stress-energy
tensor has the form $<T_{\mu\nu}>=<\rho_{vac}>g_{\mu\nu}$ which behaves
like a cosmological term with $\Lambda=8\pi G \rho_{vac}$. 

The idea that $\Lambda$ might be variable has been studied for more
than two decades (see \cite{adler,weinberg} and references therein).
In a recent paper on $\Lambda$-variability,
Overduin and Cooperstock distinguish three approaches \cite{overduin}. 
In the first approach
$\Lambda g_{\mu\nu}$ is shifted onto the right-hand side of the field
equations (1) and treated as  part of the matter content. This
approach, characterized by Overduin and Cooperstock as 
being connected to dialectic materialism, 
goes back to Gliner who interpreted $\Lambda g_{\mu\nu}$ 
as corresponding to vacuum stress-energy tensor  
with the equation of state $p=-\rho$ \cite{gliner}, to Zel'dovich 
who connected $\Lambda$ with the gravitational interaction of virtual 
particles \cite{zeldovich}, and to Linde who suggested that $\Lambda$ 
can vary \cite{andrei}. In the ref. \cite{us75}
a cosmological model was proposed with the equation of state varying
from $p=-\rho$ to $p=\rho/3$. In contrast, idealistic approach prefers 
to keep $\Lambda$ on the left-hand side of the Eq.(1) and treat it as a
constant of nature. The third approach, allowing
$\Lambda$ to vary while keeping it on the left-hand side as a geometrical
entity, was first applied by Dolgov in a model in which
a classically unstable scalar field, non-minimally coupled to gravity,
develops a negative energy density cancelling the initial positive
value of a cosmological constant $\Lambda$ \cite{dolgov}.

Whenever variability of $\Lambda$ is possible, it requires
the presence of some matter source other than
$T_{\mu\nu}=(8\pi G)^{-1}\Lambda g_{\mu\nu}$, since the conservation
equation $G^{\mu\nu}_{~~;\nu}=0$ implies $\Lambda=const$ in this case.
This requirement makes it impossible
to introduce a cosmological term as
variable in itself. 
However, it is possible for a stress-energy tensor
other than $\Lambda g_{\mu\nu}$.

The aim of this letter is to show 
what the algebraic structure of a cosmological
term can be in the spherically symmetric case, as suggested by the Petrov 
classification scheme \cite{petrov} and by the Einstein field 
equations.

In the spherically symmetric static case a line element can be
written in the form \cite{tolman}
$$
ds^2 = e^{\mu(r)}dt^2 - e^{\nu(r)} dr^2 - r^2 d\Omega^2\eqno(2)$$
where $d\Omega^2$ is the line element on the unit sphere.
The Einstein equations are 
$$ 8\pi G T_t^t = e^{-\nu}({\nu}^{\prime}/r-1/{r^2})
+{1}/{r^2}\eqno(3)$$
$$8\pi G T_r^r = -e^{-\nu} ({\mu}^{\prime}/r+{1}/{r^2})
+{1}/{r^2}\eqno(4)$$
$$8\pi G T_{\theta}^{\theta}=8\pi G T_{\phi}^{\phi}=$$
$$-e^{-\nu}({{\mu}^{\prime\prime}}/{2}+{{{\mu}^{\prime}}^2}/4
+({{\mu}^{\prime}-{\nu}^{\prime}})/{2r}-{{\mu}^{\prime}
{\nu}^{\prime}}/{4})\eqno(5)$$
A prime denotes differentiation with respect to $r$.\\
In the case of 
$$T_{\mu\nu}=(8\pi G)^{-1} \Lambda g_{\mu\nu} = \rho_{vac} g_{\mu\nu}
\eqno(6)$$
the solution is the de Sitter geometry with constant positive curvature
$R=4\Lambda$. The line element is
$$ds^2=\biggl(1-\frac{\Lambda r^2}{3}\biggr)dt^2-
\biggl({1-\frac{\Lambda r^2}{3}}\biggr)^{-1}dr^2-r^2 d\Omega^2
\eqno(7)$$
The algebraic structure of the stress-energy tensor (6), corresponding to 
a cosmological term $\Lambda g_{\mu\nu}$, is
$$ ~T_t^t=T_r^r=T_{\theta}^{\theta}=T_{\phi}^{\phi}, \eqno(8)$$
and the equation of state is
$$p=-\rho\eqno(9)$$

In the Petrov classification scheme \cite{petrov}
stress-energy tensors are
classified on the basis of their algebraic structure.
When the elementary divisors of the matrix 
$T_{\mu\nu}-\lambda g_{\mu\nu}$
(i.e., the eigenvalues of $T_{\mu\nu}$) are real, the eigenvectors of
$T_{\mu\nu}$ are nonisotropic and form a comoving reference
frame. Its timelike vector represents a velocity. The classification
of the possible 
algebraic structures of stress-energy tensors satisfying the above
conditions contains five 
possible types: [IIII],[I(III)], [II(II)], [(II)(II)], [(IIII)].
 The first symbol denotes the eigenvalue related to the timelike eigenvector.
Parentheses combine equal (degenerate) eigenvalues.
A comoving reference frame is defined uniquely
 if and only if none of the
spacelike eigenvalues ${\lambda}_{\alpha}(\alpha=1,2,3)$  coincides with
a timelike eigenvalue ${\lambda}_0$. Otherwise there exists an infinite set
of comoving reference frames. 

In this scheme the de Sitter
stress-energy tensor (6) is represented by [(IIII)] (all eigenvalues 
being equal)
and classified as a vacuum tensor due to the absence of a preferred comoving
reference frame. 
An observer moving through the de Sitter vacuum (6) cannot in principle
measure his  velocity with respect to it, since his
comoving reference frame is also comoving for (6) \cite{gliner}.

In the spherically symmetric case it is possible, by the
same definition, to introduce  an $r$-dependent vacuum
stress-energy tensor with the algebraic structure \cite{me92} 
$$T_t^t=T_r^r;~~T_{\theta}^{\theta}=T_{\phi}^{\phi}\eqno(10)$$
In the Petrov classification scheme this stress-energy tensor 
is denoted by [(II)(II)].
It has an infinite set
of comoving reference frames, since it is invariant under rotations
in the $(r,t)$ plane. Therefore an observer moving through it
cannot in principle measure the radial component of his velocity.  
The stress-energy tensor (10)
describes a spherically symmetric anisotropic vacuum 
invariant under the boosts in the radial direction \cite{me92}.

The conservation equation $T^{\mu\nu}_{~~;\nu}=0$ gives the $r$-dependent
equation of state \cite{werner,me92}
$$p_r=-\rho;~~p_{\perp}=p_r+(r/ 2)(dp_r/{dr})
\eqno(11)$$
where $\rho=T_t^t$ is the density, $p_r=-T_r^r$ is the radial
pressure, and $p_{\perp}=-T_{\theta}^{\theta}=-T_{\phi}^{\phi}$
is the tangential pressure. \\
In this case  equations (3)-(4) reduce  to the equation
$$8\pi G\rho=e^{-\nu(r)}({{\nu}^{\prime}}/{r}-{1}/{r^2})+{1}/{r^2}\eqno(12)
$$
whose solution is
$$g_{00}=e^{-\nu(r)}=1-\frac{2G{\cal M}(r)}{r};~~{\cal M}(r)
=\int_0^r{\rho(x)x^2dx}\eqno(13)$$
and the line element is 
$$ds^2=(1-{2G{\cal M}(r)}/{r})dt^2-
(1-{2G{\cal M}(r)}/{r})^{-1}dr^2-r^2d\Omega^2
\eqno(14)$$
If we require the density $\rho(r)$ to vanish as $r\rightarrow
{\infty}$ quicker then $r^{-3}$, then the metric (14) for large $r$ 
has the Schwarzschild form 
$$g_{00}(r)=1-{2GM}/{r}\eqno(15)$$
with
$$M=4\pi\int_0^{\infty}{\rho(r)r^2dr} < \infty\eqno(16)$$
If we impose the boundary condition of de Sitter behaviour (7)
at $r\rightarrow 0$, the form of the mass function
${\cal M}(r)$ in the limit of small $r$ must be \cite{werner,valera,me96}
$${\cal M}(r)=({\Lambda}/{6G})r^3=({4\pi}/{3})\rho_{vac}r^3\eqno(17)$$
For any density profile satisfying conditions (16)-(17),
the metric (14) describes a globally regular de Sitter-Schwarzschild geometry,
asymptotically Schwarzschild as $r\rightarrow{\infty}$ and
asymptotically de Sitter as $r\rightarrow 0$ \cite{me96,me99}.

The fundamental difference from the Schwarzschild case is that 
there are two horizons,
a black hole horizon $r_{+}$ and an internal Cauchy horizon $r_{-}$
\cite{werner,valera,me92}.
A critical value of the mass $M_{crit}$ exists, at which the horizons come
together. This gives a lower limit for
the black hole mass.

Depending on the value of the mass $M$, there exist
three types of configurations 
in which a Schwarzschild singularity is replaced with $\Lambda$ core
\cite{me96,me99}:
1) A $\Lambda$ black hole ($\Lambda$BH) for $M>M_{crit}$ \cite{idea};
2) An extreme $\Lambda$BH for $M=M_{crit}$; 
3) A "$\Lambda$ particle" ($\Lambda$P) - a particle-like structure without 
horizons "made up" of a self-gravitating spherically symmetric 
vacuum (10) - for $M < M_{crit}$.

In the course of Hawking evaporation, a $\Lambda$BH loses its mass
and the configuration evolves towards a $\Lambda$P \cite{me96}. 

De Sitter-Schwarzschild configurations are plotted in Fig.1  for the case
of the density profile \cite{me92,me96}
$$
\rho(r)=8\pi G\Lambda \exp{\biggl(-\frac{\Lambda}{6GM}r^3\biggr)}=
\rho_{vac} \exp{\biggl(-\frac{4\pi}{3}\frac{\rho_{vac}}{M}r^3\biggr)}
\eqno(18)$$
The mass function in the metric (14) then takes the form
$$
{\cal M}(r)=M\biggl(1-\exp{\biggl(-\frac{\Lambda}{6GM}r^3\biggr)}\biggr)\eqno(19)$$
% FIGURE 1
\begin{figure}
\vspace{-8.0mm}
\begin{center}
\epsfig{file=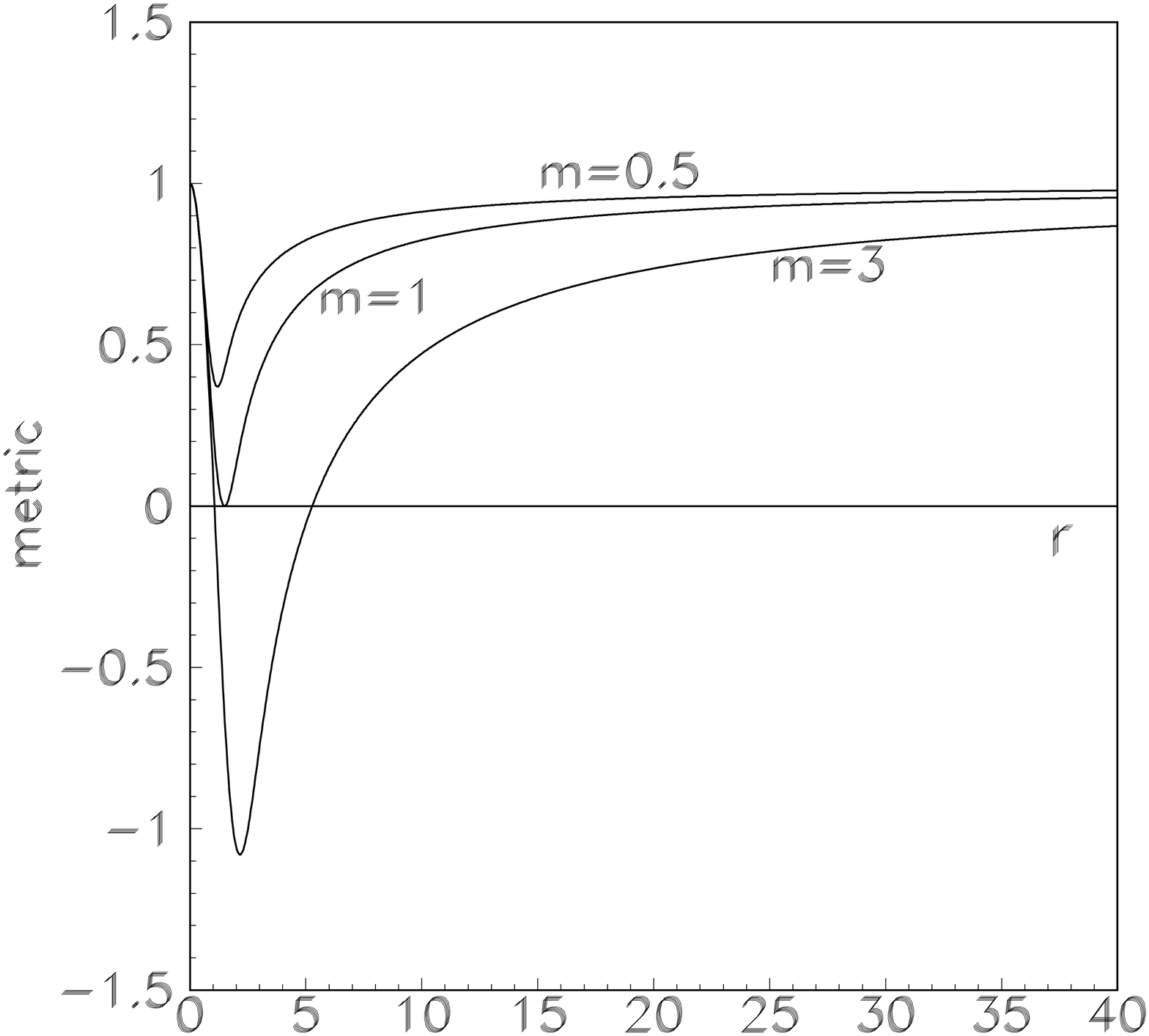,width=8.0cm,height=6.5cm}
\end{center}
\caption{
The metric coefficient $g_{00}(r)$ for de Sitter-Schwarzschild 
configurations in the case of the density profile (18). 
The parameter $m$ is the mass $M$ normalized to 
$M_{crit}\simeq{0.3M_{Pl}({\rho}_{Pl}/{\rho}_{\Lambda})^{1/2}}$.}
\label{fig.1}
\end{figure}
The stress-energy tensor (10) responsible for the $\Lambda$BH 
and $\Lambda$P solutions
connects in a smooth way two vacuum states:
de Sitter vacuum (6) at the origin and Minkowski vacuum $T_{\mu\nu}=0$
at infinity. The vacuum equation of state
(9) remains valid for the radial component of a pressure.
This makes it possible to treat the stress-energy tensor (10) as
corresponding to  an $r$-dependent  cosmological term 
$\Lambda_{\mu\nu}$, varying from $\Lambda_{\mu\nu}=\Lambda g_{\mu\nu}$ 
as $r\rightarrow 0$ to $\Lambda_{\mu\nu}=0$ as $r\rightarrow\infty$, and 
satisfying the equation of state (11) with $\rho^{\Lambda}=\Lambda^t_t$,
$p_r^{\Lambda}=-\Lambda^r_r$ and 
$p_{\perp}^{\Lambda}=-\Lambda^{\theta}_{\theta}=-\Lambda^{\phi}_{\phi}$. 

The global structure of de Sitter-Schwarzschild spacetime in the case
$M>M_{crit}$ is shown in Fig.2 \cite{me96}. It contains an infinite
sequence of $\Lambda$ black holes (${\cal BH}$), $\Lambda$ white holes
(${\cal WH}$), past and future $\Lambda$ cores (${\cal RC}$), 
and asymptotically flat universes (${\cal U}$).
A $\Lambda$ white hole models a
non-singular cosmology with inflationary origin
followed by anisotropic Kasner-type expansion due to the anisotropy
of the $\Lambda$ tensor, which in this model is time-dependent \cite{us99}.
% FIGURE 2
\begin{figure}
\vspace{-8.0mm}
\begin{center}
\epsfig{file=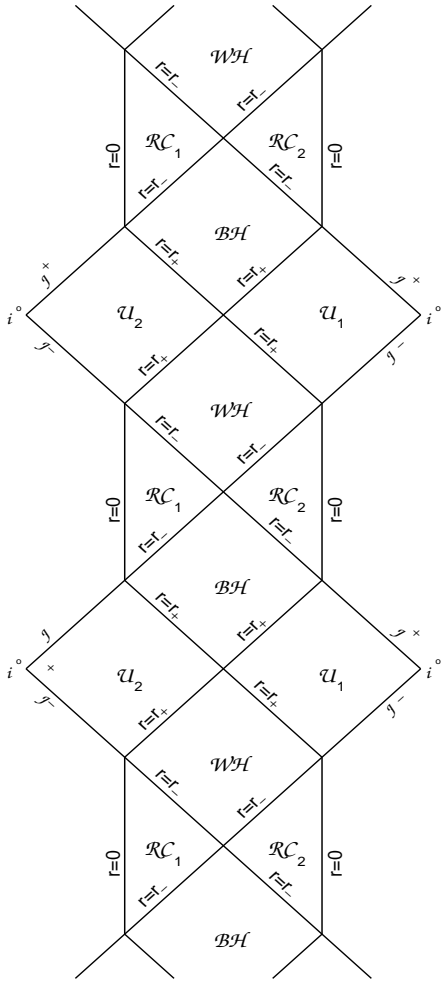,width=8.6cm,height=14.7cm} 
\end{center}
\caption{
Penrose-Carter diagram for $\Lambda$ black hole.}
\label{fig.2}
\end{figure}
The conformal diagram shown in Fig.2 represents the global structure
of de Sitter-Schwarzschild spacetime in the case of any smooth density
profile $\rho(r)$ satisfying conditions (16)-(17). In the 
case of discontinuous density profile $\rho(r)={\rho}_{Pl}\Theta(r_m-r)$ 
corresponding to direct matching of de Sitter to Schwarzschild 
metric at the junction surface $r=r_m$ \cite{werner,valera},
the asymptotically flat regions are connected 
through a black hole interior to a baby universe arising inside 
a black hole \cite{valera}.
The question of arising a baby universe inside a black hole in the
case of an arbitrary continuous density profile is considered
in the ref.\cite{us99}.

The vacuum energy outside a $\Lambda$BH horizon is given by
$$
E_{vac}=\int_{r_{+}}^{\infty}{\rho(r)r^2dr}=
M\exp{\biggl(-\frac{\Lambda}{6GM}r_{+}^3\biggr)}
\eqno(20)$$
One can say that a $\Lambda$ black hole has $\Lambda$ hair.

The question of the stability of a $\Lambda$BH and $\Lambda$P is currently
under investigation. Comparison 
of the ADM mass (16) with the proper mass \cite{MTW}
$$
{\mu}=4\pi \int_0^{\infty}{{\rho(r)}
(1-{2G{\cal M}(r)}/{r})^{-1/2}r^2dr}\eqno(21)
$$
makes a suggestion. In the spherically symmetric situations
the ADM mass represents the total energy, 
$M={\mu}+binding~ energy$ \cite{MTW}. 
In our case ${\mu}$ is bigger than $M$. This gives us a hint 
that the configuration might be stable since energy is needed 
to break it up.

If we modify the density profile to allow a non-zero value
of cosmological constant $\lambda$ as $r\rightarrow {\infty}$, putting
$$
T_t^t(r)=\rho(r)+8\pi G\lambda,
\eqno(22)$$
we obtain the metric \cite{us97}
$$
ds^2=(1-{2G{\cal M}(r)}/{r}-{\lambda r^2}/{3})dt^2- $$
$$
({1-{2G{\cal M}(r)}/{r}-{\lambda r^2}/{3}})^{-1} dr^2-r^2d\Omega^2
\eqno(23)
$$
whose asymptotics are the de Sitter metric (7) with
$(\Lambda +\lambda)$ as $r\rightarrow 0$ and with $\lambda$ as
$r\rightarrow{\infty}$.
The two-lambda spacetime has in general three horizons: a cosmological
horizon $r_{++}$, a black hole horizon $r_{+}$ and a Cauchy
horizon $r_{-}$. 
Horizons are calculated by solving the equation $g_{00}(r)=0$
with $g_{00}(r)$ from the Eq.(23).
They are plotted in Fig.3 for the case of the
density profile given by (18). There are two
critical values of the mass $M$, restricting the BH mass from
below and above. 
A lower limit $M_{cr1}$ corresponds to the first extreme BH
state $r_{+}=r_{-}$
and is very close to the lower limit for $\Lambda$BH.
An upper limit $M_{cr2}$ corresponds to the second extreme state
$r_{+}=r_{++}$ and depends on the parameter 
$q\equiv{\sqrt{{\Lambda}/{\lambda}}}$. 
% FIGURE 3
\begin{figure}
\vspace{-8.0mm}
\begin{center}
\epsfig{file=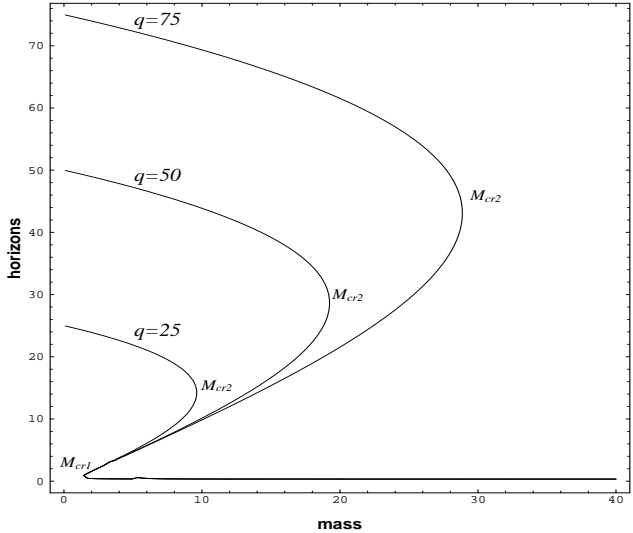,width=9.0cm,height=7.5cm}
\end{center}
\caption{
Horizon-mass diagram forn a two-lambda spacetime.
An upper limit for a BH mass $M_{cr2}$ depends on the parameter
$q\equiv{(\Lambda/\lambda)^{1/2}}$.}
\label{fig.3}
\end{figure}
Depending on the mass $M$, two-lambda geometries represent
five types of configurations \cite{us98}:\\
1) A two-lambda black hole ($\Lambda\lambda$BH) for $M_{cr1}<M<M_{cr2}$,
which is a non-singular cosmological black hole, i.e. a non-singular 
modification of the Kottler-Trefftz solution \cite{kot}, 
frequently referred to in the literature as 
a Schwarzschild-de Sitter black hole.
2) An extreme $\Lambda\lambda$BH with the minimum possible mass 
$M=M_{cr1}$.
3)A $\Lambda$P with a de Sitter background of small $\lambda$ for
$M<M_{cr1}$.
4) An extreme $\Lambda\lambda$BH with the maximum possible mass 
$M=M_{cr2}$,
which is the non-singular modification of the Nariai solution \cite{nariai}.
5) Soliton-like configuration for $M>M_{cr2}$, a one-horizon
solution with lambda  varying from $\Lambda$ at the origin
to $\lambda$ at infinity, which can be called a "$\Lambda$-bag".
These configurations are plotted in Fig.4.
% FIGURE 4
\begin{figure}
\vspace{-8.0mm}
\begin{center}
\epsfig{file=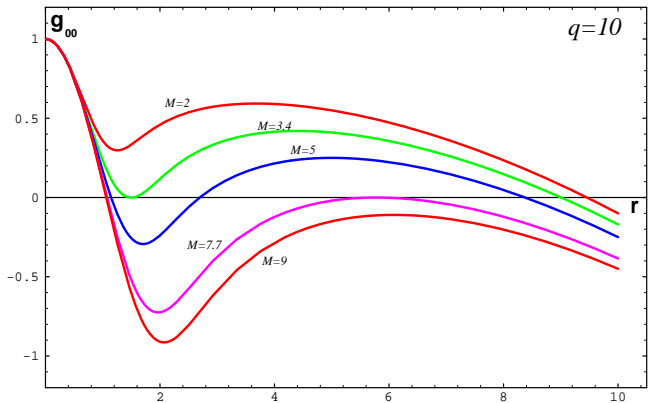,width=9.0cm,height=6.0cm} 
\end{center}
\caption{
Two-lambda configurations for the case $q=10$.
The mass $M$ is normalized to $(3/G^2 \Lambda )^{1/2}$.
Two extreme states for $\Lambda\lambda$BH are $M_{cr1}\simeq{3.4}$ 
and $M_{cr2}\simeq{7.7}$.}
\label{fig.4}
\end{figure}
The global structure of two-lambda spacetimes 
for the case $M_{cr1}<M<M_{cr2}$ is shown in Fig.5.
It contains an infinite sequence of $\Lambda\lambda$BH, $\Lambda\lambda$WH
(white holes with $\Lambda$ at the origin and $\lambda$ at infinity), 
$\Lambda$ future and past cores, and asymptotically
de Sitter universes  with small $\lambda$.
A two-lambda white hole models non-singular
cosmology with inflationary origin followed by anisotropic Kasner-like
stage and ended in $\lambda$ dominated stage \cite{us2000}.
% FIGURE 5
\begin{figure}
\vspace{-8.0mm}
\begin{center}
\epsfig{file=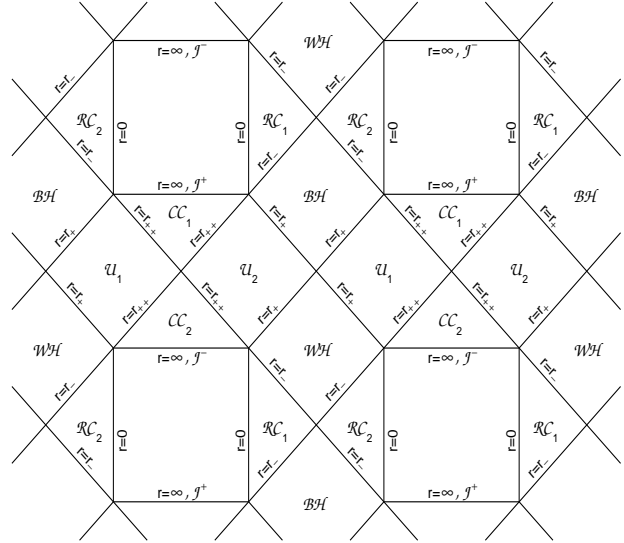,width=8.5cm,height=7.5cm}
\end{center}
\caption{
Penrose-Carter diagram for $\Lambda\lambda$ black holes.
There is an infinite sequence
of black and white holes ${\cal BH}$,${\cal WH}$, whose singularities are
replaced by $\Lambda$ cores ${\cal RC}_1,{\cal RC}_2$, asymptotically
de Sitter universes ${\cal U}_1,{\cal U}_2$, and $\lambda$ cores 
${\cal CC}_1, {\cal CC}_2$
(regions beyond the cosmological horizons $r_{++}$).}
\label{fig.5}
\end{figure}
The stress-energy tensor responsible for the two-lambda geometry connects
in a smooth way two vacuum states with non-zero cosmological constant:
de Sitter vacuum $T_{\mu\nu}=(8\pi G)^{-1} (\Lambda+\lambda)g_{\mu\nu}$
at the origin, and de Sitter vacuum $T_{\mu\nu}=(8\pi G)^{-1}\lambda
g_{\mu\nu}$ at infinity. This confirms the proposed interpretation
of the stress-energy tensor (10) as corresponding to a variable 
effective cosmological term
$\Lambda_{\mu\nu}$.

In conclusion, let us compare the proposed variable cosmological tensor
$\Lambda_{\mu\nu}$ with the quintessence which is a time-varying,
spatially inhomogeneous component of matter content with negative
pressure \cite{CDS}. The key difference comes from the equation of state.
For quintessence the equation of state is $p=-\alpha\rho$ with
$\alpha < 1$ \cite{CDS}. This corresponds to such a 
stress-energy tensor $T_{\mu\nu}$ for which a comoving
reference frame is defined uniquely. The quintessence represents
thus a non-vacuum alternative to a cosmological constant $\Lambda$,
while the tensor $\Lambda_{\mu\nu}$ represents the extension of 
the algebraic structure of a cosmological term 
which allows it to be variable. 
\vskip0.2in
{\bf Acknowledgement}

This work was supported by the Polish Committee for Scientific Research
through the Grant 2.P03D.017.11.

% start references ++++

% stop references


\begin{references}
\bibitem{idea} The idea goes back to Gliner paper \cite{gliner}
where he suggested that de Sitter vacuum 
can be a final state in a gravitational collapse.
\end{references}
\end{document}